# Proton Pencil-Beam Scanning Stereotactic Body Radiation Therapy and Hypofractionated Radiation Therapy for Thoracic Malignancies: Patterns of Practice Survey and Recommendations for Future Development from NRG Oncology and PTCOG

Wei Liu, PhD[1*], Hongying Feng, PhD[1*], Paige A. Taylor, MS[2], Minglei Kang, PhD[3], Jiajian Shen, PhD[1], Jatinder Saini, PhD[4], Jun Zhou, PhD[5], Huan B. Giap, MD, PhD[6], Nathan Y. Yu, MD[1], Terence S. Sio, MD[1], Pranshu Mohindra, MD[7], Joe Y. Chang, MD[8], Jeffrey D. Bradley, MD[9], Ying Xiao, PhD[9], Charles B. Simone II, MD[3], Liyong Lin, PhD[5]

[1]Department of Radiation Oncology, Mayo Clinic, Phoenix, AZ 85054, USA

[2]The Imaging and Radiation Oncology Core Houston Quality Assurance Center, The University of Texas MD Anderson Cancer Center, Houston, Texas.

[3]New York Proton Center, New York City, New York, USA

[4]Seattle Cancer Care Alliance Proton Therapy Center and Department of Radiation Oncology, University of Washington School of Medicine, Seattle, WA 98195, USA.

[5]Department of Radiation Oncology and Winship Cancer Institute, Emory University, Atlanta, Georgia

[6]Department of Radiation Oncology, Medical University of South Carolina, Charleston, South Carolina, USA

[7]Department of Radiation Oncology, University Hospitals Cleveland Medical Center, Cleveland, Ohio, USA

[8]Department of Radiation Oncology, University of Texas MD Anderson Cancer Center, Houston

[9]Department of Radiation Oncology, University of Pennsylvania, Philadelphia, Pennsylvania, USA

*Co-first authors who contribute to this paper equally

Corresponding author: Wei Liu, Liu.Wei@mayo.edu.

Author responsible for statistical analysis: Wei Liu, Liu.Wei@mayo.edu.

**Funding Statement**

This research was supported by the National Cancer Institute (NCI) Career Developmental Award K25CA168984, President's Discovery Translational Program of Mayo Clinic, the Fred C. and Katherine B. Andersen Foundation Translational Cancer Research Award, Arizona Biomedical Research Commission Investigator Award, the Lawrence W. and Marilyn W. Matteson Fund for Cancer Research, and the Kemper Marley Foundation.

**Data Availability Statement**

Research data are stored in an institutional repository and will be shared upon request to the corresponding author.

**Abstract**

Stereotactic body radiation therapy (SBRT) and hypofractionation using pencil-beam scanning (PBS) proton therapy (PBSPT) is an attractive option for thoracic malignancies. Combining the advantages of target coverage conformity and critical organ sparing from both PBSPT and SBRT, this new delivery technique has great potential to improve the therapeutic ratio, particularly for tumors near critical organs. Safe and effective implementation of PBSPT SBRT/hypofractionation to treat thoracic malignancies is more challenging than the conventionally-fractionated PBSPT due to concerns of amplified uncertainties at the larger dose per fraction. NRG Oncology and Particle Therapy Cooperative Group (PTCOG) Thoracic Subcommittee surveyed US proton centers to identify practice patterns of thoracic PBSPT SBRT/hypofractionation. From these patterns, we present recommendations for future technical development of proton SBRT/hypofractionation for thoracic treatment. Amongst other points, the recommendations highlight the need for volumetric image guidance and multiple CT-based robust optimization and robustness tools to minimize further the impact of uncertainties associated with respiratory motion. Advances in direct motion analysis techniques are urgently needed to supplement current motion management techniques.

**Introduction**

Lung cancer is the leading cause of cancer-related deaths and one of the most common malignancies diagnosed in the United States and worldwide[1,2]., particularly for medically inoperable patients.

Radiation therapy is a standard treatment option for thoracic malignancies[3]. The conventional prescription for the treatment of thoracic malignancies is 1.8 to 2 Gy per fraction as once-daily course. Accelerated hyperfractionation is employed for small cell lung cancer patients treated with 1.5 Gy per fraction twice daily. While similar hyperfractionation techniques with lower (1.1-1.2 Gy per fraction) were also explored as part of chemoradiation, no clear advantage was established in the setting of lung tumors. In contrast, advances in fundamental radiotherapy technologies including image guidance, treatment planning, and treatment delivery have allowed safe delivery of stereotactic body radiotherapy (SBRT[4]/SABR[5]; 7-34 Gy per fraction in 1-8 fractions (per the clinical trial PACIFIC-4 NCT03833154 and SWOG S1914 NCT04214262)) and hypofractionated regimens (2.5-8 Gy per fraction) at much higher dose-per fraction. Compared to conventional radiotherapy that relies on the differential radiation repair between tumor and normal tissue, SBRT seeks to "ablate" tumors while limiting normal tissue irradiation with highly conformal plans that rely on accurate delivery[6]. Hypofractionation is an intermediate between conventional fractionation and SBRT.

SBRT or SABR can be delivered by either photon or proton beam therapy. Numerous photon-based SBRT studies for stage I/II non-small cell lung cancer have demonstrated excellent outcomes[7-10]. Consensus guidelines describing the optimal utilization and technical requirements for SBRT implementation with photons when treating thoracic malignancies have been published[4,11]. Compared to photon therapy, the intrinsic Bragg peak sparing characteristic of proton therapy that deposits all of its energy within a certain range, makes it a feasible option for SBRT or SABR in lung cancers[12].

Due to the physical characteristics of proton dose deposition (e.g., sharp distal fall-off), proton dose distributions are more sensitive to various forms of uncertainties than photon dose distributions[13-58]. First, proton range uncertainty due to the conversion of the Hounsfield unit (HU) to proton stopping power for dose calculation can degrade the dose distribution quality[13], i.e., the altered proton range could result in either underdose in the target or overdose in the adjacent organs at risk (OARs). Second, the tissue density heterogeneity in the thorax region (airways in the trachea and bronchus, low-density lung parenchyma, and high-density vessels, liver, bone, and heart) can have a significant impact on dose distributions, especially for pencil beam dose calculation algorithms. Though having high calculation efficiency, pencil beam dose calculation algorithms cannot achieve the desired dose calculation accuracy in heterogeneous media due to poor ability to account for the inhomogeneities[52,53,55-59]. Third, intra-fractional motion (primarily from breathing) is the biggest challenge in implementing proton for thoracic malignancies. The motion perturbs the locations of the tumor and normal tissues, which directly blurs the dose gradient between targets and nearby normal tissues, thus downgrading target coverage and normal tissue sparing[14]. Compared to the more "static" passive scattering proton delivery technique to generate broad beams by double scattering, the pencil beam scanning (PBS) proton delivery technique is more "dynamic" where the whole target is covered by a series of scanning proton beamlets (several millimeters in size) controlled by two orthogonal magnetic field in a certain order. Unlike the nature of broad beam in passive scattering proton therapy, the PBS dose distribution is determined by the dose distribution of each pristine proton beamlet, which is very sensitive to various uncertainties. Therefore, for PBS in particular, the interplay effect[17,22,26,34,37,43,60-64] caused by the interference between dynamic beamlet delivery and intra-fractional tumor motion can cause additional degradation of the delivered dose distribution. Higher fraction dose and Limited fraction number in proton SBRT/hypofractionation can compound above uncertainties because of the absence of the averaging effect of multiple fractions in a conventionally fractionated treatment[65].

As such, motion management (during planning and treatment delivery) is critical for accurate delivery of proton therapy for thoracic malignancies, esp. with SBRT/ hypofractionation approach. Each institution should perform institution-specific measurements, assessments, and management, due to inter-institutional variations in infrastructure (hardware and software), for example, machine-specific spot size and monitor unit (MU) limit and availability of different motion monitoring and analysis tools.

Several proton centers in the United States have begun to implement proton SBRT/hypofractionation for lung cancer. In 2021, a questionnaire on the current clinical practice of proton SBRT and hypofractionation in the treatment of lung cancers was distributed to all US proton centers participating in NRG Oncology clinical trials. The questionnaire (included in Supplemental Materials) was designed by the NRG Oncology Working Group on Proton Lung SBRT/Hypofractionation and consisted of 83 questions, covering 7 categories: (1) Vendors, Delivery Techniques, and Treatment Planning System (TPS), (2) Patient Selection, (3) Simulation and immobilization, (4) Treatment Planning and Quality Assurance (QA), (5) Image-Guided Radiation Therapy (IGRT) and Motion Management, (6) Follow-up, and (7) General Questions. Most questions were of multiple selection type, with several explicitly given candidate answers (commonly seen in clinical practice) and one "Other" option for less common or institution-specific responses. In addition, every question allowed comments for further explanation and/or customization.

The current study, on behalf of the NRG Oncology Proton SBRT/Hypofractionation Working Group based on the questionnaire, presents a comprehensive description of the current status (clinical workflow and approaches) of implementing proton SBRT/hypofractionation for lung cancers, and discusses current needs to further optimize the proton SBRT/hypofractionation practice for thoracic malignancies.

**Response overview**

The survey was distributed to 30 proton centers in the United States, among which 24 proton centers in the US responded to the questionnaire (response rate 80%). Table A1 lists the distributions of proton system vendors, beam delivery techniques, and treatment planning systems (TPS). Among the 24 proton centers, 11 had clinical practices in place for treating thoracic malignancies using proton SBRT and/or hypofractionation, and 10 proton centers completed the corresponding Imaging and Radiation Oncology Core (IROC) proton lung phantom credentialing. One proton center expressed its intention to implement proton SBRT/hypofractionation in the treatment of lung cancer in the near future. Notably, the quantitative evaluations of the questionnaire hereafter are not necessarily restricted to the responses from the proton centers that currently practiced proton SBRT/hypofractionation in lung cancer treatment (11 proton centers from the survey). One proton center that had clinical proton SBRT practices employed double scattering delivery technology, thus excluded in this quantitative evaluation focused on PBS. Another two proton centers that had not performed proton SBRT/hypofractionation for lung cancer yet, provided insightful information and responses on this topic, thus included in the quantitative evaluation.

For most of the questions, there was a consensus response. A wide variety of responses, however, was identified for a few questions, which were then categorized into two perspective groups based on the related topics: physicists and physicians (see Table 1).

Table 1. Questions that received a variety of responses from two perspectives of physicists or physicians.

| Physicists | Physicians |
|---|---|
| **1. Patient CT Simulation**<br>a. What CT scan slice thickness is typically used for lung SBRT/hypofractionation?<br>**2. Treatment Planning**<br>a. How many beams are typically used?<br>b. What is the dose volume criterion for target evaluation for robustness analysis?<br>**3. Motion Management and IGRT**<br>a. What motion management methods are used?<br>b. What equipment is used for DIBH?<br>c. What rescanning techniques is used (in-layer or volumetric)?<br>d. What types of IGRT are used?<br>**4. General**<br>a. What are the 3 most important technologies to improve lung treatment quality? | **1. Patient Selection**<br>a. What is the maximum tumor diameter allowed?<br>b. What is the prescription for primary lung cancers?<br>c. What are the dose constraints for uninvolved ipsilateral and contralateral lung?<br>**2. Treatment Planning**<br>a. What are the dosimetry parameters for the following targets (PTV, CTV, and GTV)?<br>b. Which target is used for coverage evaluation for the nominal plan?<br>c. What is the inhomogeneity dose allowed in the target?<br>d. What is the allowable mean lung-GTV dose for 5 fx SBRT (if applicable)?<br>**3. Follow up**<br>a. What is the process for patient follow-up?<br>**4. General**<br>a. How can we increase the use of proton SBRT in lung cancer treatment? |

**Patient Selection**

The use of curative proton SBRT/hypofractionation was restricted to patients with an ECOG score ≤2 for performance status (8/8[1]), early cancer stages (5/7 for stage I and 2/7 for stage I and II), and ≤ 3 discrete lesions (10/10). To treat patients with inoperable stage I/II lung cancers, institutions usually adopt proton SBRT/hypofractionation (9/10). Proton SBRT/hypofractionation was also used in re-irradiation (11/11) and lung metastases (9/11). If proton SBRT/hypofractionation was used to treat patients with implanted cardiac devices (7/11), it was important to ensure that the patients were not critically dependent on the device (6/7), sometimes with additional constraints such as maximum dose to the device less than 2 Gy [RBE] and regular device interrogation.

Several centers (4/11) set no limits for the maximum tumor diameter allowed, while other centers had institution-specific limits (Figure 1 (a)). The prescribed fractionation was also diverse (Figure 1 (b)), not only inter-institutionally but also intra-institutionally, primarily based on different tumor location. For instance, one center prescribed 48 Gy [RBE] in 4 fractions for most patients, while 50 Gy [RBE] in 5 fractions for centrally located tumors. Another center prescribed 60 Gy [RBE] in 5 fractions or 54 Gy [RBE] in 3 fractions for peripherally located tumors, but 50

---

[1] Hereafter, the numerator is the frequency of each effective answer from the survey, whereas the denominator is the number of proton centers that provided the certain effective answers. Notably, for some questions, certain proton centers provided multiple choices, which resulted in the numerators from not adding up to the denominators. The denominator sometimes could also be less than 12 because: 1) a few proton centers of the 12 proton centers included in the data analysis did not provided answers to the questions and were thus neglected; 2) the question was a conditional one that only a few qualified proton centers were required to provide answers.

Gy [RBE] in 5 fractions for centrally located tumors. Due to the difference in prescribed fractionation, the dose volume constraints (DVCs) for the uninvolved ipsilateral and contralateral lung were also varied accordingly (Table 2).

Table 2. Constraints for uninvolved ipsilateral and contralateral lung from different proton centers that provided effective responses[*].

| # | Prescription | Constraints | Comment |
|---|---|---|---|
| 1 | 50Gy/5fx (central), 48Gy/4fx (most) | $V_{20Gy} \leq 10\%$ | |
| 2 | 50Gy/5fx | $V_{13.5Gy} \leq 37\%$, $V_{12.5Gy} \leq 1500cc$, $V_{13.5Gy} \leq 1000cc$ | For both lungs, per RTOG 0813 |
| 3 | 50Gy/5fx | $D_{mean} < 8Gy[RBE]$, $V_{20Gy} < 10\% \sim 15\%$ | Per HyTEC for SBRT[66], no real constraint for hypofractionation |
| 4[a] | 60Gy/15fx | $V_{13Gy} \leq 35\%$, $V_{7Gy} \leq 40\%$, $V_{3Gy} \leq 60\%$ | |
| 5 | 50Gy/5fx | $V_{15Gy} \leq 1000cc$, $V_{12Gy} \leq 1500cc$ | |
| 6 | 60Gy/5fx or 54Gy/3fx (peripherally), 50Gy/5fx (central) | $V_{13Gy} \leq 36\%$, | |
| 7 | 50Gy/5fx | $V_{12.5Gy} \leq 1500cc$, $V_{13.5Gy} \leq 1000cc$ | Avoid shooting toward/through the contralateral lung; use multiple beams with smaller separation |
| 8[a] | 50~60Gy in 15~20 fx | $D_{mean} < 18 Gy[RBE]$, $V_{20Gy} < 30\%$, $V_{10Gy} < 40\%$ | Per PCG LUN005 study |
| 9 | 50Gy/5fx | $D_{mean} < 6Gy[RBE]$ (total lung) $D_{mean} < 10 Gy[RBE]$ (ipsilateral lung) | As low as possible for contralateral lung |

[*] Constraints for proton SBRT only. For photon SBRT constrains please see the reference[67].

[a] Hypofractionation only

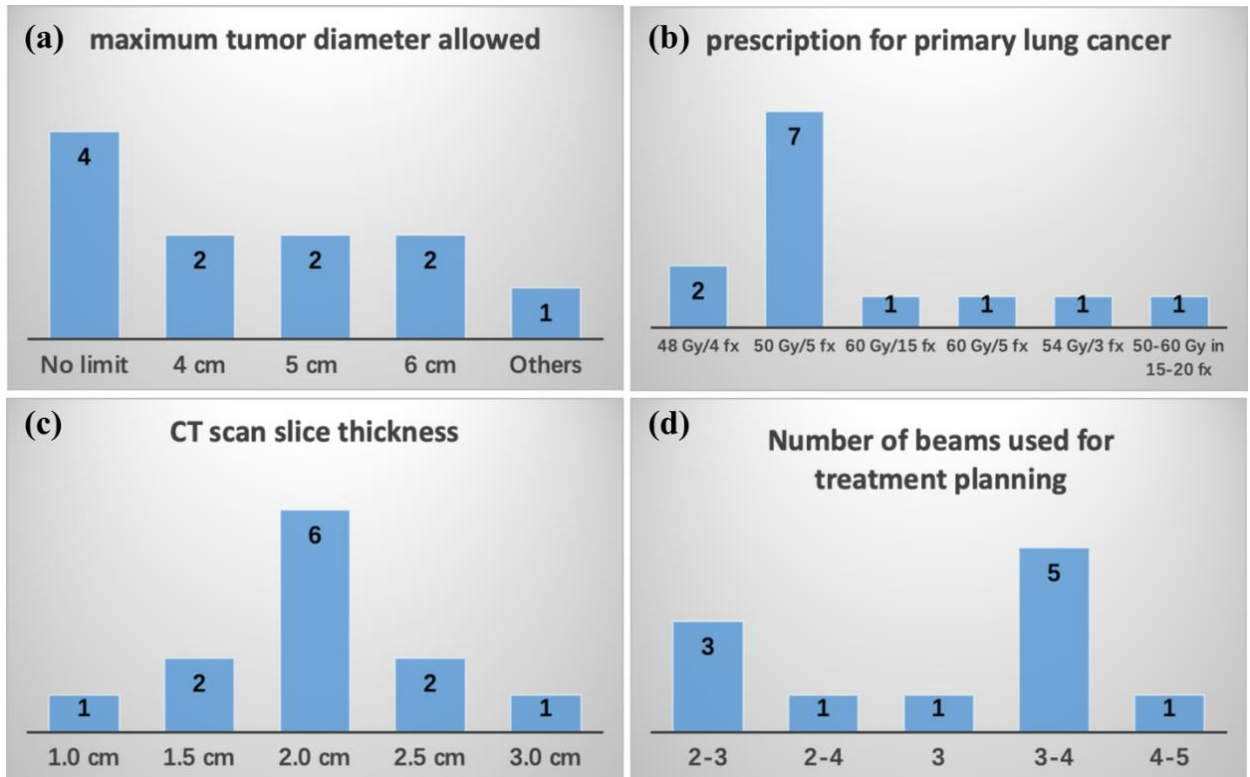

Figure 1. The summarized counts of (a) limit of maximum tumor diameter allowed, (b) prescription for primary lung cancer, (c) CT scan slice thickness, and (d) number of beams used for treatment planning from different proton centers.

**Patient Immobilization**

VacLoc was the dominant method for patient immobilization (10/12), while other methods included the use of thermoplastic body mask (2/12), green foam (2/12), and CIVCO and Klarity (1/12); one center used all 4 methods. Abdominal compression, which could be considered a patient immobilization method, is discussed in the Motion Management section.

**Treatment Simulation, Contouring, and Target Definition**

Four-dimensional (4D) CT was routinely used for diagnosis and motion analysis (discussed in detail in the Motion Management section) in proton SBRT/hypofractionation for the treatment of lung cancer (12/12). While 4D CTs was sorted mainly using phase-based method(11/12), one proton center used amplitude-based method. Different slice thicknesses (Figure 1 (c)) was used in the CT scan, with 2.0 mm being the most used (6/12). After determining the motion magnitude with the patient immobilization techniques previously discussed, a threshold of motion magnitude was usually chosen (11/12) to initialize certain motion management techniques (discussed in detail in the Motion Management section). The most frequently used threshold was 5 mm (6/12), whereas a less stringent threshold of 1 cm (5/12) was also reported. When the motion magnitude was below these thresholds, patients could be treated with free breathing, otherwise, patients would be re-simulated with either breath-hold (BH) or compression belt. For deep-inspiration breath hold (DIBH), simulations were repeated several times (3, 5, or 7 times) to quantify residual tumor motion during DIBH CTs.

After the acquisition of multiple CTs (4D or multiple DIBH), a 3D CT scan was typically generated for target delineation, which was usually the maximum intensity projection (MIP) scan (7/12) or averaged scan (3/12). To assist with the target delineation, the patients received PET/CT scans (11/11) and MRI scans (3/11). The NRG protocols (8/10) and institution-specific internal protocols (3/10) were used for target and OAR delineation.

Once the gross tumor volume (GTV) was identified, some centers (2/12) directly adopted it as the treatment target, while other centers (10/12) created a clinical target volume (CTV) by adding an expansion to the GTV to account for the sub-clinical extension of the tumor. The expansion was usually isotropic with a 5 mm margin (4/7; 3 centers did not provide details of their expansion) or defined according to physician preference (3/7). Several OARs were cropped out of the CTV expansion always or sometimes: heart (4/7 (always) and 2/7(sometimes)), chest wall (3/7 and 2/7), spinal cord (6/7 and 1/7), and esophagus (5/7 and 2/7). Internal target volume (ITV) was another target volume routinely used in PBSPT for the treatment of tumors with motion, like lung cancers, and was typically created by (1) adding another margin to CTV to compensate for the motion or (2) adding another margin to internal GTV (IGTV) to account for microscopic disease. Usually, IGTV was formed either by contouring in the MIP CT images or by combining all GTVs of every phase of a 4D CT (or multiple DIBH CT) images. Sometimes (5/10), a further expansion, usually 5mm isotropically (5/5), from CTV/ITV was added to create a planning target volume (PTV) for reporting and evaluation purposes. Notably, the PTV concept has limitations and thus is not suitable to be used in robust optimization for PBSPT[68].

**Treatment Planning**

To obtain the optimal tumoricidal dose for targets and protection for OARs, various treatment beams were selected per institution (Figure 1 (d)). In-house tools was developed and used by one proton center to facilitate the search for optimal beam angles[69].

The plans were optimized through a single-field optimization (SFO) technique (5/11), where target dose homogeneity was retained within each field, or a multi-field optimization (MFO) technique (4/11), where target dose homogeneity was retained given all fields combined. Compared to SFO, MFO has higher flexibility in optimization, thus often making it easier to meet the dosimetric requirements but also more sensitive to target motion and uncertainties. Therefore,

MFO was often used as an alternative option to SFO (3/4) for challenging cases. A hybrid plan (6/11) could also be generated by adding two weighted plans that used SFO and MFO, or a controlled SFO with higher beam dose.

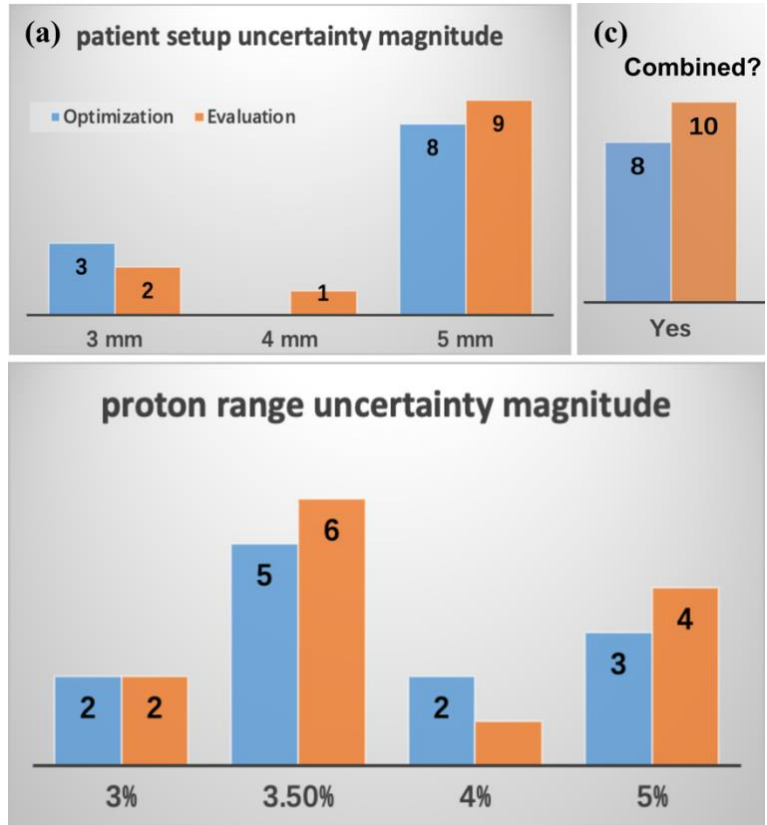

Figure 2. Summary of different robust optimization and robustness evaluation configurations: (a) patient setup uncertainty magnitude, (b) proton range uncertainty, and (c) whether combining patient setup and proton range uncertainties, in robust optimization and robustness evaluation respectively, of different proton centers.

Robust optimization[15,18,19,21,25,26,31,33,34,41,42,68,70-73] was the routine method for treatment planning of stage I/II lung cancer using proton SBRT/hypofractionation, in which patient setup and proton range uncertainties were explicitly included in the optimization algorithm (11/12), with the exceptional one center because only SDX (DYN'R, France) images were used in the treatment planning. Besides targets, dose distributions on the adjacent OARs were sometimes robustly optimized in the treatment planning as well in most centers (10/12). The worst-case (9/12) and second worst (2/12) scenario was the dominant strategy for robust optimization. The patient setup uncertainty was usually addressed by an isotropic rigid shift in all possible cardinal directions with a certain magnitude (Figure 2(a)), whereas the proton range uncertainty was considered by scaling the relative stopping power (RSP) in the CT image up and down by a certain percentage (Figure 2(b)). However, patient setup and proton range uncertainties were only sometimes simultaneously taken into account (simultaneously included in one perturbed scenario in the robust optimization) (8/11, Figure 2(c)). Combinations of patient setup and proton range uncertainties (when simultaneously considered) were ±3 mm/±3.5% (1/8), ±5 mm/±3% (1/8), ±5 mm/±3.5% (2/8), ±5 mm/±4% (3/8), and ±5 mm/±5% (3/8).

Table 3. Dose volume constraints used by different proton centers the provided effective responses for the treatment planning of proton SBRT/hypofractionation in stage I/II lung cancer[*].

| # | PTV | CTV | GTV |
|---|---|---|---|
| 1 | $D_{99\%} \geq 90\%$[a] | $V_{110\%} < 20\%$[b] | Not used |
| 2 | $V_{100\%} > 95\%$[c] | Not used | Not used |
| 3 | Not used | $V_{100\%} > 95\%$, $D_{99\%} > 90\%$ | Not used |
| 4 | $D_{95\%} \geq 95\%$ | $V_{95\%} > 95\%$ | -[d] |
| 5 | - | $D_{95\%} \geq 100\%$ | $D_{95\%} \geq 100\%$ |
| 6 | Not used | $D_{98\%} > 100\%$ | $D_{98\%} > 100\%$ |
| 7 | Not used | $D_{99\%} \geq 100\%$ | Not used |
| 8 | - | $D_{95\%} \geq 100\%$ | $D_{98\%} \geq 100\%$ |
| 9 | - | $D_{95\%} \geq 100\%$ | $D_{95\%} \geq 100\%$ |
| 10 | $D_{95\%} \geq 95\%$ | $D_{95\%} \geq 100\%$ | $D_{99\%} > 100\%$ |
| 11 | - | $D_{95\%} \geq 100\%$ | - |

[*]Constraints for proton SBRT only. For photon SBRT constrains please see the reference[67].

[a] $D_{x\%} \geq y\%$, at least $x\%$ of the structure volume receives dose no less $y\%$ of the prescription dose.

[b] $V_{x\%} < y\%$, the volume receives at least $x\%$ of the prescription dose is less than $y\%$ of the structure.

[c] $V_{x\%} > y\%$, the volume receives at least $x\%$ of the prescription dose is larger than $y\%$ of the structure.

[d] No answer provided.

To address respiratory-induced tumor motion in proton SBRT/hypofractionation treatment planning, the ITV (7/10) was usually the optimization target for 3D robust optimization on a 3D averaged CT (9/11) with density overrides of soft tissue (8/9). For different targets (GTV, CTV, and PTV), different dose volume constraints (DVCs) were used for robust optimization, listed in Table 3.

Amongst available dose calculation engines[54-56,58] for robust treatment planning, analytical (4/12) and Monte Carlo (MC) dose calculation algorithms (8/12) are used in clinical practices. Typically dose grid resolutions were 2mm (7/12), 2.5mm (4/12), and 3mm (1/12). If MC algorithms were used, the statistical uncertainty in the target regions was kept below 0.5% (7/8).

**Image Guidance and Motion Management**

Image guidance is critical for patient alignment and accurate beam delivery, and this is often of greater significance when using proton SBRT/hypofractionation to treat stage I/II lung cancer, due to the small volume of the target and reduced averaging effect with fewer fractions (Figure 1(a)(b)). The most common image-guidance technique was projected x-ray based, from the basic 2 dimensional (2D) orthogonal kV image (9/12). Newer proton centers utilized space-resolved faster but blurrier cone-beam CT (CBCT) (7/12) or slower but diagnostically equivalent CT-on-rails (CToR) (3/12). Individual proton centers were equipped with multiple image-guided modalities and used them solely (orthogonal kV/CBCT/CToR, 6/12), in simultaneous combination (orthogonal kV and CBCT, 5/12), or in asynchronous complementarity (orthogonal kV and CToR, 2/12). During patient alignment, the position of the patient (guiding images) was matched to the planning images by assessing the mutual deviations of structures, including the target (12/12) and other regions of interest (ROIs) (7/12). The ROIs could be isodose line contours (3/7) and other bony/landmark structures (4/7), e.g., spinal cord, carina, and diaphragm. The deviation tolerance was 3 mm (7/11) or 2 mm (4/11).

If motion management was required at a particular stage of the treatment course, several approaches were used (Table A2). All the motion management techniques could be used solely or together, depending on the patient's characteristics, conditions, and techniques available at the treating institution.

DIBH (9/12) could be used in both treatment simulation and treatment delivery to mitigate the impact of respiratory motion and better protect OARs. For centers equipped with CBCT, a half-scan was usually used during DIBH (5/7). Equipment to obtain the respiratory motion information to guide DIBH was commercially available, including Active Breath Coordinator$^{TM}$ (Eletka, Stockholm, Sweden) (1/9), Real-time Position Management$^{TM}$ (Varian, Palo Alto, CA) (2/7), SDX (5/9) and alignRT (visionRT, London, UK) (1/9). Abdominal compression is, in principle, similar to DIBH to reduce tumor motion by using a compression plate/belt/corset on the abdomen to limit the magnitude of patient respiration without respiratory training. Anzai (Anzai Medical Co., Ltd., Tokyo, Japan) abdominal pressure sensor was also used for respiratory motion analysis to guide abdominal compression (1/9). All the time-resolved respiratory information (lung volume, body surface, and abdominal pressure) acquired by the aforementioned respiratory monitoring systems and fluoroscopy (x-ray videos) could also be used to do real-time tumor tracking with the help of fiducial markers. Using the aforementioned respiratory monitoring systems, gating was also used to mitigate tumor motion impact (3/12) in treatment delivery. By defining a gating window (usually selecting certain respiratory phases associated with 4D CTs), the beam delivery could be synchronized with the tumor/surrogate motion. Rescanning[74-76], also known as repainting, was another simple technique used to minimize the dose uncertainty caused by the interplay effect in the treatment delivery (10/12) by visiting each spot position multiple times. Due to the attenuated averaging effect of decreased fraction number in proton SBRT/hypofractionation, the averaging effect brought by rescanning became more indispensable. Rescanning could be layered (4/10) and volumetric (7/10). In layered rescanning, each energy layer was rescanned entirely before proceeding to the next energy layer, while in volumetric rescanning, the whole target volume was repetitively scanned. Range shifter (RS) with a larger air gap was used to produce a larger spot size (10/12), which would result in improved dose homogeneity in moving targets[17].

**Plan quality evaluation**

After the plan is optimized, a comprehensive plan quality evaluation should be completed. The evaluation usually consisted of an independent second dose calculation (10/11), linear energy transfer (LET) evaluation (3/11), and robustness evaluation (12/12) on the planning CT and a few CTs associated with different respiratory phases. For the second dose calculation, analytical (4/10) and MC (6/10) dose engines were used per institution. For the patients treated with free breathing, the 2 extreme breathing phase CTs (maximum inhale and maximum exhale) were usually used for the plan evaluation besides the planning CT. Similarly, for the patients undergoing DIBH or gating, other independent DIBH scans or phase CT scans (within the selected gating window) could be used for the plan evaluation besides the planning CT. The original plan could only be approved to treat the patient until all of the required target coverage, and normal tissue criteria are met on all CTs, otherwise, the original plan should be adjusted.

The plan quality evaluation matrices included target coverage (12/12) (with varied targets per institution, Figure 3(a)), conformity index (CI) (5/12) (with varied inhomogeneity dose allowed per institution, Figure 3(b)), and non-involved lung protection (refer to the constraints in

Table 2). The LET evaluation is a relatively new metric that has yet to have a consensus on its necessity or the LET-related constraints that could be referred to. Therefore, only in-house developed tools have been utilized, either based on analytical (1/3) or MC (3/3) algorithms. The robustness configurations in plan robust evaluation differed slightly from those in plan robust optimization (Figure 2). The evaluating matrices were also varied (Figure 3 (c)) based on either the dose-volume histogram (DVH) indices in the worst-case scenario or the bandwidths of DVH indices from the DVH family for all uncertainty scenarios.

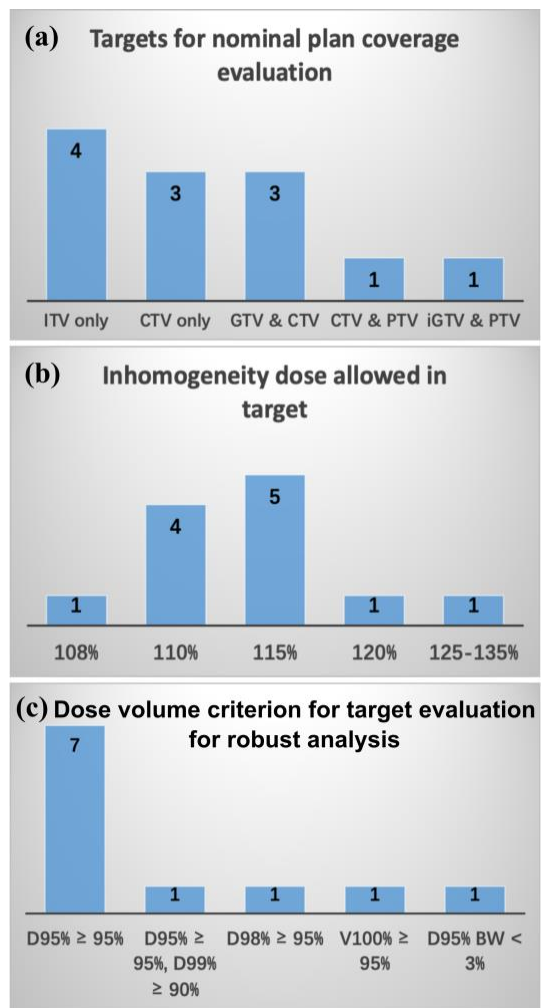

Figure 3. Summary of different (a) targets for nominal plan coverage evaluation, (b) inhomogeneity dose allowed in targets, and (c) dose volume criterion for target evaluation for robust analysis ,from different proton centers. BW is the bandwidth of DVH indices from the DVH family for all uncertainty scenarios.

The interplay effect was also evaluated (3/12) either by measurement (1/3) or by calculation (2/3). In-house software to evaluate the interplay effect was developed by calculating the 4D dynamic dose (4DDD)[20,22,24,26-28,34,37,43,45,46,60-63,77]. To calculate 4DDD, time-dependent spot delivery was considered together with the time-dependent changes in anatomy. First, every spot for each field per fraction was assigned to the corresponding respiratory phase according to the temporal relationship between the spot delivery sequence and patient-specific respiratory motions. Then, the dose of each phase was calculated for the assigned spots only. Finally, the calculated dose of each phase was then deformed to the reference phase through deformable image registration (DIR)[78] to get the final 4DDD.

The plan measurement-based patient-specific quality assurance (PSQA) verifies what is delivered against what has been planned/calculated (11/11). This is typically done via Gamma analysis[79,80] that quantifies the agreement between the calculated dose distribution and the measured dose distribution, in terms of distance to agreement (DTA, mm) and dose difference (DD, %). The dose calculation resolution for PSQA was usually 2 mm (8/11), and 2.5 mm (2/11). With pre-selected criteria for both DTA and DD, a threshold of the Gamma analysis passing rate was set to determine the acceptance (i.e., exceeding the threshold) of the plan, for example, a threshold of 90% of pixels with the criteria of 2mm/2% (3/10), a threshold of 90% with the criteria of 3mm/3% (6/10), and a threshold of 95% with the criteria of 3mm/3% (1/10).

**Follow-up**

The general follow-up strategy includes follow-up imaging (PET/CT or CT) and history and physical (H&P) exams with a gradual reduction in frequency (Table 4). While CT was always used, PET/CT was sometimes used 3-6 months post-radiation or for suspected recurrence.

Table 4. Follow-up process and the imaging modality from different proton centers that provided effective responses.

| # | Phase 1* | | Phase 2 | | Phase 3 | |
|---|---|---|---|---|---|---|
| | Image modality | Frequency and Duration | Image modality | Frequency and Duration | Image modality | Frequency and Duration |
| 1 | CT[a] | Q6 mon, 2 yr | CT[a] | Annual, --[b] | [c] | |
| 2 | CT | Q3 mon, -- | | | | |
| 3 | PET/CT | Q3 mon, 2 yr | | | | |
| 4 | CT | NA, 1 yr | | | | |
| 5 | PET/CT | Q3 mon, NA | PET/CT | Q3-6 mon, -- | | |
| 6 | PET/CT[d] | Q3-4 mon, 2 yr | CT | Q4-6 mon, 3 yr | CT | Annual, -- |
| 7 | PET/CT | Q3-4 mon, 1 yr | PET/CT | SOF, -- | | |

*abbreviations*: NA for Not Answered, SOF for Spreading Out Frequency, Q*x* mon for every (quaque) *x* months, *x* yr for *x* years.

* Follow-up phase is decided based on the follow-up frequency and the corresponding duration.

[a] PET was used when suspected progression was diagnosed.

[b] Continue until the death of the patient or when the patient drops from the follow-up.

[c] No follow-up.

[d] PET/CT usually for the first post-radiation follow-up

**Barriers to Implementation of Proton Lung SBRT/Hypofractionation**

For the two general questions for physicists and physicians, we received 20 effective responses (Figure 4), which shed light on unmet clinical needs. From the physicians' perspective (Figure 4(a)), the lack of technologies is the major barrier to more widespread utilization of proton SBRT in lung cancer treatment. This includes reducing range uncertainty, increasing conformity for small targets, improving robust optimization using Monte Carlo, improving CBCT quality, volumetric or layer repainting, tumor tracking, and proton arc. Education and dissemination of knowledge and clinical evidence to support proton SBRT in lung cancer treatment was the second most mentioned problem that needs future work. This is despite a meta-analysis reporting improved outcomes and reduced toxicities when delivering proton hypofractionation relative to photon SBRT[81], and an increasing number of prospective reports of safe and effective proton SBRT delivery[82]. Insurance approval was also a significant barrier. From the perspective of physicists who were asked to name the 3 most important technologies to improve lung treatment quality (Figure 4(b)), volumetric IGRT ranked first. Motion management and fast treatment were the second most mentioned. Online adaptive planning, robust optimization, and improved gating/breath holding integration were also considerably important. Some other issues, e.g., incorporation of MC dose engine in treatment planning, automated repainting, real-time tumor tracking was identified by a few participants as well.

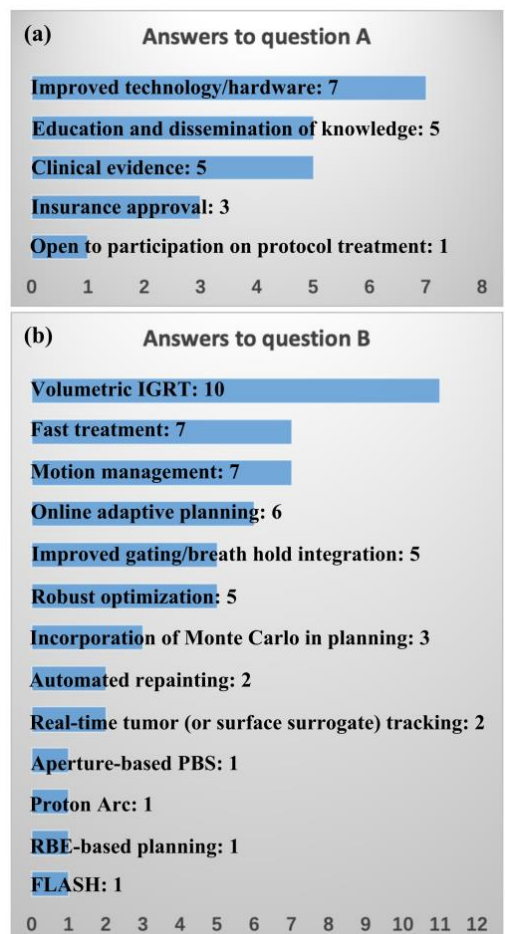

Figure 4. Answers to question (a) How can we increase the use of proton SBRT in lung cancer treatment? (b) What are the 3 most important technologies to improve lung treatment quality?

**Recommended Technical Implementations and Future Developments**

SBRT has been considered as a standard treatment approach for early-stage lung cancer[11,83] and increasingly been used for intrathoracic oligometastatic or oligoprogressive disease[84] to achieve durable disease control. In medically-inoperable early-stage non-small cell lung cancer (NSCLC), photon-based SBRT resulted in better local control and overall survival for smaller, none-negative peripheral tumors, compared to conventionally fractionated radiation therapy[85]. Higher rates of local control were observed in photon-based SBRT for both peripheral and central NSCLC compared to moderate hyperfractionation[86]. Although photon-based SBRT has shown great promise across tumor stages and lung cancer histologies, there is reluctance to fully embrace it across the full spectrum of lung cancer, based on well-founded toxicity concerns such as in larger tumors[81,87] and in central and ultra-central tumors[10,88,89]. Compared to photons, the unique physical properties of protons (especially delivered in PBS fashion) allow them to deliver tumoricidal doses of irradiation while limiting the volume of normal tissue exposed to a low dose bath, thus leading to improved overall survival and reduced rates of toxicity in lung cancers[48,90,91]. Therefore, PBSPT-based SBRT has emerged as a solution to these toxicity risks and challenges with photon-based SBRT. Several proton centers in the United States have begun to implement proton SBRT/hypofractionation for lung cancer, with more to be expected thanks to the reported encouraging results[81,92] with early data, including randomized data and meta-analysis data, supporting the benefits of protons over photons in early-stage NSCLC, and with the increased accessibility of PBSPT, especially in those newly constructed proton centers. Under such circumstances, a national survey was conducted to report the practice patterns of proton SBRT/hypofractionation in thoracic malignancies with a goal of identifying points of consensus and areas for future improvement.

Fundamental inter-institutional differences in target delineation were shown in the survey. For instance, methods to delineate the targets from GTV to ITV differed considerably. When phase CT in a multiple-CT set (4DCT or multiple DIBH CTs) was used for target delineation, per International Commission on Radiation Units & Measurements (ICRU) Report 78, GTVs on each CT were expanded by a margin to form CTVs to account for sub-clinical extension of the tumor, which were then expanded by another margin to form an ITV to address internal uncertainties (e.g., motions). The order in addressing the internal uncertainties and sub-clinical disease could alternatively be switched and margin amplitudes also varied from physician to physician and from institution to institution. When a MIP CT was used for target delineation, the motion induced internal uncertainties were inherently included by the MIP CT images, thus the GTV delineated on MIP CT images was essentially internal GTV (IGTV). Such differences could lead to different target DVCs in both plan optimization and evaluation. The margins used in the definitions of different targets try to balance the tumoricidal dose coverage and the protection of adjacent normal tissues in the presence of all kinds of uncertainties. The potent fraction dose of SBRT calls for more attention to the protection of adjacent normal tissues. Therefore, the margins used in the target definitions should preferably be smaller than those used in conventionally fractionated proton therapy, which further calls for more accurate image guidance, uncertainty consideration, and motion management. At the same time, a guiding protocol is needed to regulate the target definition process further.

For IGRT technologies, CToR can provide high-quality volumetric imaging, but take longer operation time and usually result in more diagnostic dose to patients. CBCT can also provide volumetric IGRT but suffers from low image quality. Algorithms have been proposed to enhance the quality of CBCT images[93-96]. The 4D CBCT has also been developed for better image guidance for mobile tumors[97-100]. In addition, improvements in CBCT can promote its use in daily online adaptive therapy[23,101-104].

Motion management is another major area that needs technology improvements. In the robust optimization stage, 3D robust optimization with ITV as the target is the current clinical routine. Centers equipped with the more recent research version of RayStation or dedicatedly developed in-house tools had the capability of conducting the 4D robust optimization[22,37,105-109]. In the 4D robust optimization, multiple CT images from different phases (theoretically all but typically representative) with CTVs (phase-specific delineation or copied from the corresponding 3D CT) as the main targets were incorporated to explicitly consider the respiratory motion in the optimization algorithm through the 4D accumulated dose (4DD) on the reference CT or uniform dose distributions on all the selected CTs, negating the need for an ITV. The 4DD on the reference CT was the phase-averaged sum of the dose calculated on all incorporated phases and warped to the reference CT via DIR. For simplicity, the two extreme respiratory phases (maximum inhalation and exhalation) could be used in the 4D robust optimization[22,37]. Note that the idea of 4D robust optimization based on 4DCTs could be analogously translated to the robust optimization using multiple DIBH CTs or multiple phase CTs[110] in the gating window. In the stage of plan evaluation, the interplay effect should also be evaluated since newly built proton centers are equipped with PBS-based proton machines. Unfortunately, like the multiple CT (4D CTs, or multiple DIBH CTs) robust optimization, the interplay evaluation is only supported by a few in-house developed tools developed[43,111-113] and the more recently research version of one of the vendors (RayStation). Therefore, we recommend that proton centers engage vendors to provide to options of multiple CT robust optimization and interplay effect evaluation. In the stages of treatment simulation and treatment delivery, current technologies are performing indirect analysis of tumor motion, i.e., surrogate-based, for instance, lung volume, abdomen surface, abdominal pressure, etc. However, the movement of surrogates is not necessarily synchronized with the tumor motion. Fluoroscopy may help give internal anatomical information, yet only in 2D. Better motion analysis can be achieved by (1) building a model that reflects the comprehensive relationship between tumor motion and the motion of all the surrogates and (2) developing easily-invasive or non-invasive techniques for direct motion analysis. The advances in motion analysis will further improve motion management based on it, e.g., DIBH, abdominal compression, and gating. In addition, fast treatment delivery and automated repainting from the hardware aspect is also desired for better motion management.

MC methods are recognized to be more accurate than analytical algorithms for proton dose calculation, especially in heterogeneous media, e.g., thoracic area. However, the time consumption prevents the clinical use of MC with robust optimization due to intensive computation needs. Recently, MC dose engines with simplified physics models and graphic processing unit (GPU) acceleration have enabled the dose calculation of a typical proton plan within minutes, and even seconds[57,114-121]. With the state-of-the-art super-fast MC dose engine, the MC-based robust optimization (even multiple CTs-based) can be conducted, which will prominently improve the dose accuracy in proton SBRT for lung cancer.

FLASH effect[122] where, with ultrahigh dose rate (UHDR, >40Gy/s) and high dose, normal tissue radiosensitivity is significantly reduced whereas tumor control is maintained potentially allows a novel radiation therapy modality to further increase the therapeutic ratio. A recent study[123] exploited the feasibility of proton FLASH-SBRT, using patient-specific ridge filters to spread the Bragg peak from a fixed transmitting energy beam to a proximal beam-specific planning target volume.

Artificial intelligence (AI)-based technology is the new boost to the radiotherapy community in all aspects, let alone its applications people can expect for proton SBRT in lung cancer treatment. Auto-segmentation of targets and organs is currently the most acknowledged application of AI in radiotherapy. AI-based DIR[124-126], the fundamental technology in calculating 4DD and 4DDD, has also been reported with equivalent or even better accuracy but dramatically increased efficiency compared to the conventional DIR algorithms. AI-based denoising technologies in MC dose engines can vastly improve the MC simulation efficiency with acceptable accuracy[127,128]. AI-based dose engines have also been developed and can achieve MC simulation equivalent accuracy within seconds for calculating a typical proton plan dose[129]. We believe incorporating AI-based technologies as an alternative or complementary method will further promote using proton SBRT for lung cancer.

With the development of technologies, additional proton SBRT clinical trials need to be carried out. By collecting the patient outcomes from clinical trials, proton centers need to (1) provide evidence on the efficacy of proton SBRT in lung cancer treatment, (2) gather experiences and establish protocols for a standard workflow of proton SBRT in lung cancer treatment and each component within, (3) convince physicians, physicists, and insurance providers of the superior effectiveness of SBRT in treating lung cancer, and (4) educate physicians and physicists to conduct standardized proton SBRT for lung cancers.

This report showed the ongoing clinical use of proton SBRT for lung cancers. In this report, we have tried to report how proton SBRT for lung cancers had been performed from different proton centers to identify the challenges and to highlight areas in need of additional research and development. Proton SBRT is quite complex and technology-intensive. No one report can include all the needed information and provide omniscient instructions. Enrollment in a clinical trial is encouraged for proton SBRT in lung cancer, as proton SBRT for non-small lung cancer is a component of the recently activated NRG Oncology trial LU008, and we hope this report could serve as a stepping stone for development of that and future clinical trial.

# Reference

1. Siegel RL, Miller KD, Wagle NS, Jemal A. Cancer statistics, 2023. *CA: A Cancer Journal for Clinicians.* 2023;73(1):17-48.
2. ACS. Cancer Facts & Figures 2022. In.: American Cancer Society Atlanta, GA, USA; 2022.
3. Simone CBI, Rengan R. The Use of Proton Therapy in the Treatment of Lung Cancers. *The Cancer Journal.* 2014;20(6):427-432.
4. Potters L, Kavanagh B, Galvin JM, et al. American Society for Therapeutic Radiology and Oncology (ASTRO) and American College of Radiology (ACR) Practice Guideline for the Performance of Stereotactic Body Radiation Therapy. *International Journal of Radiation Oncology, Biology, Physics.* 2010;76(2):326-332.
5. Loo BW, Jr., Chang JY, Dawson LA, et al. Stereotactic ablative radiotherapy: what's in a name? *Practical Radiation Oncology.* 2011;1(1):38-39.
6. Timmerman RD, Herman J, Cho LC. Emergence of stereotactic body radiation therapy and its impact on current and future clinical practice. *Journal of Clinical Oncology.* 2014;32(26):2847.
7. Timmerman R, Paulus R, Galvin J, et al. Stereotactic body radiation therapy for inoperable early stage lung cancer. *Jama.* 2010;303(11):1070-1076.
8. Bezjak A, Paulus R, Gaspar LE, et al. Safety and Efficacy of a Five-Fraction Stereotactic Body Radiotherapy Schedule for Centrally Located Non-Small-Cell Lung Cancer: NRG Oncology/RTOG 0813 Trial [published online ahead of print 20190403]. *J Clin Oncol.* 2019;37(15):1316-1325.
9. Modh A, Rimner A, Williams E, et al. Local control and toxicity in a large cohort of central lung tumors treated with stereotactic body radiation therapy. *International Journal of Radiation Oncology* Biology* Physics.* 2014;90(5):1168-1176.
10. Roach MC, Robinson CG, DeWees TA, et al. Stereotactic Body Radiation Therapy for Central Early-Stage NSCLC: Results of a Prospective Phase I/II Trial. *Journal of Thoracic Oncology.* 2018;13(11):1727-1732.
11. Videtic GMM, Donington J, Giuliani M, et al. Stereotactic body radiation therapy for early-stage non-small cell lung cancer: Executive Summary of an ASTRO Evidence-Based Guideline. *Practical Radiation Oncology.* 2017;7(5):295-301.
12. Diwanji T, Sawant A, Sio TT, Patel NV, Mohindra P. Proton stereotactic body radiation therapy for non-small cell lung cancer. *Annals of Translational Medicine.* 2020;8(18):1198.
13. Lomax AJ. Intensity modulated proton therapy and its sensitivity to treatment uncertainties 1: the potential effects of calculational uncertainties. *Physics in Medicine and Biology.* 2008;53(4):1027-1042.
14. Lomax AJ. Intensity modulated proton therapy and its sensitivity to treatment uncertainties 2: the potential effects of inter-fraction and inter-field motions. *Physics in Medicine and Biology.* 2008;53(4):1043-1056.
15. Liu W, Zhang X, Li Y, Mohan R. Robust optimization in intensity-modulated proton therapy. *Med Phys.* 2012;39:1079-1091.
16. Stuschke M, Kaiser A, Pottgen C, Lubcke W, Farr J. Potentials of robust intensity modulated scanning proton plans for locally advanced lung cancer in comparison to intensity modulated photon plans. *Radiotherapy and oncology : journal of the European Society for Therapeutic Radiology and Oncology.* 2012;104(1):45-51.
17. Grassberger C, Dowdell S, Lomax A, et al. Motion Interplay as a Function of Patient Parameters and Spot Size in Spot Scanning Proton Therapy for Lung Cancer. *International Journal of Radiation Oncology Biology Physics.* 2013;86(2):380-386.

18. An Y, Liang JM, Schild SE, Bues M, Liu W. Robust treatment planning with conditional value at risk chance constraints in intensity- modulated proton therapy. *Medical Physics.* 2017;44(1):28-36.
19. An Y, Shan J, Patel SH, et al. Robust intensity-modulated proton therapy to reduce high linear energy transfer in organs at risk. *Medical Physics.* 2017;44(12):6138-6147.
20. Feng H, Sio TT, Rule WG, et al. Beam angle comparison for distal esophageal carcinoma patients treated with intensity-modulated proton therapy [published online ahead of print 2020/10/16]. *J Appl Clin Med Phys.* 2020;21(11):141-152.
21. Feng H, Shan J, Anderson JD, et al. Per-voxel constraints to minimize hot spots in linear energy transfer (LET)-guided robust optimization for base of skull head and neck cancer patients in IMPT. *Med Phys.* 2021.
22. Feng H, Shan J, Ashman JB, et al. Technical Note: 4D robust optimization in small spot intensity-modulated proton therapy (IMPT) for distal esophageal carcinoma [published online ahead of print 2021/06/01]. *Med Phys.* 2021. doi: 10.1002/mp.15003.
23. Feng H, Patel SH, Wong WW, et al. GPU-accelerated Monte Carlo-based online adaptive proton therapy: A feasibility study. *Medical Physics.* 2022;49(6):3550-3563.
24. Liu C, Bhangoo RS, Sio TT, et al. Dosimetric comparison of distal esophageal carcinoma plans for patients treated with small-spot intensity-modulated proton versus volumetric-modulated arc therapies [published online ahead of print 2019/05/22]. *J Appl Clin Med Phys.* 2019;20(7):15-27.
25. Liu C, Patel SH, Shan J, et al. Robust Optimization for Intensity-Modulated Proton Therapy to Redistribute High Linear Energy Transfer (LET) from Nearby Critical Organs to Tumors in Head and Neck Cancer [published online ahead of print 2020/01/29]. *Int J Radiat Oncol Biol Phys.* 2020. doi: 10.1016/j.ijrobp.2020.01.013.
26. Liu CB, Schild SE, Chang JY, et al. Impact of Spot Size and Spacing on the Quality of Robustly Optimized Intensity Modulated Proton Therapy Plans for Lung Cancer. *International Journal of Radiation Oncology Biology Physics.* 2018;101(2):479-489.
27. Liu C, Sio TT, Deng W, et al. Small-spot intensity-modulated proton therapy and volumetric-modulated arc therapies for patients with locally advanced non-small-cell lung cancer: A dosimetric comparative study [published online ahead of print 2018/10/18]. *J Appl Clin Med Phys.* 2018;19(6):140-148.
28. Liu C, Yu NY, Shan J, et al. Technical Note: Treatment planning system (TPS) approximations matter - comparing intensity-modulated proton therapy (IMPT) plan quality and robustness between a commercial and an in-house developed TPS for nonsmall cell lung cancer (NSCLC) [published online ahead of print 2019/09/10]. *Med Phys.* 2019;46(11):4755-4762.
29. Liu W, Inventor. System and Method For Robust Intensity-modulated Proton Therapy Planning. 09/02/2014, 2014.
30. Liu W, ed *Robustness quantification and robust optimization in intensity-modulated proton therapy.* Springer; 2015. Rath A, Sahoo N, eds. Particle Radiotherapy: Emerging Technology for Treatment of Cancer.
31. Liu W, Frank SJ, Li X, et al. Effectiveness of Robust Optimization in Intensity-Modulated Proton Therapy Planning for Head and Neck Cancers. *Med Phys.* 2013;40(5):051711-051718.
32. Liu W, Frank SJ, Li X, Li Y, Zhu RX, Mohan R. PTV-based IMPT optimization incorporating planning risk volumes vs robust optimization. *Medical Physics.* 2013;40(2):021709-021708.
33. Liu W, Li Y, Li X, Cao W, Zhang X. Influence of robust optimization in intensity-modulated proton therapy with different dose delivery techniques. *Med Phys.* 2012;39.
34. Liu W, Liao Z, Schild SE, et al. Impact of respiratory motion on worst-case scenario optimized intensity modulated proton therapy for lung cancers. *Practical Radiation Oncology.* 2015;5(2):e77-86.

35. Liu W, Mohan R, Park P, et al. Dosimetric benefits of robust treatment planning for intensity modulated proton therapy for base-of-skull cancers. *Practical Radiation Oncology.* 2014;4:384-391.
36. Liu W, Patel SH, Harrington DP, et al. Exploratory study of the association of volumetric modulated arc therapy (VMAT) plan robustness with local failure in head and neck cancer [published online ahead of print 2017/05/16]. *J Appl Clin Med Phys.* 2017;18(4):76-83.
37. Liu W, Schild SE, Chang JY, et al. Exploratory Study of 4D versus 3D Robust Optimization in Intensity Modulated Proton Therapy for Lung Cancer. *International Journal of Radiation Oncology Biology Physics.* 2016;95(1):523-533.
38. Matney J, Park PC, Bluett J, et al. Effects of Respiratory Motion on Passively Scattered Proton Therapy Versus Intensity Modulated Photon Therapy for Stage III Lung Cancer: Are Proton Plans More Sensitive to Breathing Motion? *International Journal of Radiation Oncology Biology Physics.* 2013;87(3):576-582.
39. Matney JE, Park PC, Li H, et al. Perturbation of water-equivalent thickness as a surrogate for respiratory motion in proton therapy [published online ahead of print 2016/04/14]. *J Appl Clin Med Phys.* 2016;17(2):5795.
40. Schild SE, Rule WG, Ashman JB, et al. Proton beam therapy for locally advanced lung cancer: A review. *World journal of clinical oncology.* 2014;5(4):568-575.
41. Shan J, An Y, Bues M, Schild SE, Liu W. Robust optimization in IMPT using quadratic objective functions to account for the minimum MU constraint. *Medical Physics.* 2018;45(1):460-469.
42. Shan J, Sio TT, Liu C, Schild SE, Bues M, Liu W. A novel and individualized robust optimization method using normalized dose interval volume constraints (NDIVC) for intensity-modulated proton radiotherapy [published online ahead of print 2018/11/06]. *Med Phys.* 2018. doi: 10.1002/mp.13276.
43. Shan J, Yang Y, Schild SE, et al. Intensity-modulated proton therapy (IMPT) interplay effect evaluation of asymmetric breathing with simultaneous uncertainty considerations in patients with non-small cell lung cancer [published online ahead of print 2020/09/24]. *Med Phys.* 2020;47(11):5428-5440.
44. Tryggestad EJ, Liu W, Pepin MD, Hallemeier CL, Sio TT. Managing treatment-related uncertainties in proton beam radiotherapy for gastrointestinal cancers. *Journal of gastrointestinal oncology.* 2020;11(1):212-224.
45. Younkin J, Bues M, Keole S, Stoker J, Shen J. Multiple Energy Extraction Reduces Beam Delivery Time for a Synchrotron-Based Proton Spot-Scanning System. *Medical Physics.* 2017;44(6):2872-2872.
46. Younkin JE, Morales DH, Shen J, et al. Technical Note: Multiple energy extraction techniques for synchrotron-based proton delivery systems may exacerbate motion interplay effects in lung cancer treatments [published online ahead of print 2021/06/27]. *Med Phys.* 2021. doi: 10.1002/mp.15056.
47. Yu NY, DeWees TA, Liu C, et al. Early Outcomes of Patients With Locally Advanced Non-small Cell Lung Cancer Treated With Intensity-Modulated Proton Therapy Versus Intensity-Modulated Radiation Therapy: The Mayo Clinic Experience. *Advances in Radiation Oncology.* doi: 10.1016/j.adro.2019.08.001.
48. Yu NY, DeWees TA, Voss MM, et al. Cardiopulmonary Toxicity Following Intensity-Modulated Proton Therapy (IMPT) Versus Intensity-Modulated Radiation Therapy (IMRT) for Stage III Non-Small Cell Lung Cancer. *Clinical Lung Cancer.* 2022;23(8):e526-e535.
49. Zhang P, Fan N, Shan J, Schild SE, Bues M, Liu W. Mixed integer programming with dose-volume constraints in intensity-modulated proton therapy [published online ahead of print 2017/07/07]. *J Appl Clin Med Phys.* 2017;18(5):29-35.

50. Zaghian M, Cao W, Liu W, et al. Comparison of linear and nonlinear programming approaches for "worst case dose" and "minmax" robust optimization of intensity-modulated proton therapy dose distributions [published online ahead of print 2017/03/17]. *J Appl Clin Med Phys.* 2017;18(2):15-25.
51. Zaghian M, Lim G, Liu W, Mohan R. An Automatic Approach for Satisfying Dose-Volume Constraints in Linear Fluence Map Optimization for IMPT [published online ahead of print 2014/12/17]. *J Cancer Ther.* 2014;5(2):198-207.
52. Taylor PA, Kry SF, Followill DS. Pencil Beam Algorithms Are Unsuitable for Proton Dose Calculations in Lung. *International Journal of Radiation Oncology*Biology*Physics.* 2017;99(3):750-756.
53. Sasidharan BK, Aljabab S, Saini J, et al. Clinical Monte Carlo versus Pencil Beam Treatment Planning in Nasopharyngeal Patients Receiving IMPT. *International Journal of Particle Therapy.* 2019;5(4):32-40.
54. Deng W, Ding X, Younkin JE, et al. Hybrid 3D analytical linear energy transfer calculation algorithm based on precalculated data from Monte Carlo simulations [published online ahead of print 2019/11/24]. *Med Phys.* 2019. doi: 10.1002/mp.13934.
55. Deng W, Younkin JE, Souris K, et al. Technical Note: Integrating an open source Monte Carlo code "MCsquare" for clinical use in intensity-modulated proton therapy [published online ahead of print 2020/03/11]. *Med Phys.* 2020. doi: 10.1002/mp.14125.
56. Holmes J, Shen J, Shan J, et al. Technical note: Evaluation and second check of a commercial Monte Carlo dose engine for small-field apertures in pencil beam scanning proton therapy. *Medical Physics.* 2022;49(5):3497-3506.
57. Shan J, Feng H, Morales DH, et al. Virtual Particle Monte Carlo (VPMC), a new concept to avoid simulating secondary particles in proton therapy dose calculation. *Medical Physics.* 2022.
58. Younkin JE, Morales DH, Shen J, et al. Clinical Validation of a Ray-Casting Analytical Dose Engine for Spot Scanning Proton Delivery Systems [published online ahead of print 2019/11/23]. *Technol Cancer Res Treat.* 2019;18:1533033819887182.
59. Saini J, Traneus E, Maes D, et al. Advanced proton beam dosimetry part I: review and performance evaluation of dose calculation algorithms. *Translational Lung Cancer Research.* 2018;7(2):171-179.
60. Li Y, Kardar L, Li X, et al. On the interplay effects with proton scanning beams in stage III lung cancer. *Medical Physics.* 2014;41(2).
61. Kardar L, Li Y, Li X, et al. Evaluation and mitigation of the interplay effects of intensity modulated proton therapy for lung cancer in a clinical setting. *Practical Radiation Oncology.* 2014;4(6):e259-268.
62. Kardar L, Li Y, Li X, et al. Evaluation and mitigation of the interplay effects of intensity modulated proton therapy for lung cancer in a clinical setting. *Practical Radiation Oncology.* 2014;4(6):e259-e268.
63. Li Y, Kardar L, Li X, et al. On the interplay effects with proton scanning beams in stage III lung cancer. *Medical Physics.* 2014;41(2):021721.
64. Kang M, Huang S, Solberg TD, et al. A study of the beam-specific interplay effect in proton pencil beam scanning delivery in lung cancer. *Acta Oncologica.* 2017;56(4):531-540.
65. Knopf A-C, Lomax AJ. In the context of radiosurgery – Pros and cons of rescanning as a solution for treating moving targets with scanned particle beams. *Physica Medica.* 2014;30(5):551-554.
66. Kong F-M, Moiseenko V, Zhao J, et al. Organs at Risk Considerations for Thoracic Stereotactic Body Radiation Therapy: What Is Safe for Lung Parenchyma? *International Journal of Radiation Oncology*Biology*Physics.* 2021;110(1):172-187.
67. Puckett LL, Titi M, Kujundzic K, et al. Consensus Quality Measures and Dose Constraints for Lung Cancer from the Veterans Affairs Radiation Oncology Quality Surveillance Program and American

Society for Radiation Oncology (ASTRO) Expert Panel. *Practical Radiation Oncology.* doi: 10.1016/j.prro.2023.04.003.
68. Unkelbach J, Alber M, Bangert M, et al. Robust radiotherapy planning. *Physics in Medicine & Biology.* 2018;63(22):22TR02.
69. Cao W, Lim GJ, Lee A, et al. Uncertainty incorporated beam angle optimization for IMPT treatment planning. *Medical Physics.* 2012;39(8):5248-5256.
70. Pflugfelder D, Wilkens JJ, Oelfke U. Worst case optimization: a method to account for uncertainties in the optimization of intensity modulated proton therapy. *Physics in Medicine and Biology.* 2008;53(6):1689-1700.
71. Unkelbach J, Bortfeld T, Martin BC, Soukup M. Reducing the sensitivity of IMPT treatment plans to setup errors and range uncertainties via probabilistic treatment planning. *Medical Physics.* 2009;36(1):149-163.
72. Fredriksson A, Forsgren A, Hardemark B. Minimax optimization for handling range and setup uncertainties in proton therapy. *Medical Physics.* 2011;38(3):1672-1684.
73. Chen W, Unkelbach J, Trofimov A, et al. Including robustness in multi-criteria optimization for intensity-modulated proton therapy. *Physics in Medicine and Biology.* 2012;57(3):591-608.
74. Phillips MH, Pedroni E, Blattmann H, Boehringer T, Coray A, Scheib S. Effects of respiration motion on dose uniformity with a charged particle scanning method. *Phys Med Biol.* 1992;37:223-233.
75. Lambert J, Suchowerska N, McKenzie DR, Jackson M. Intrafractional motion during proton beam scanning. *Physics in Medicine and Biology.* 2005;50(20):4853-4862.
76. Bernatowicz K, Lomax AJ, Knopf A. Comparative study of layered and volumetric rescanning for different scanning speeds of proton beam in liver patients. *Physics in Medicine & Biology.* 2013;58(22):7905.
77. Li H, Li Y, Zhang X, et al. Dynamically accumulated dose and 4D accumulated dose for moving tumors. *Medical Physics.* 2012;39(12):7359-7367.
78. Wang H, Dong L, O'Daniel J, et al. Validation of an accelerated 'demons' algorithm for deformable image registration in radiation therapy. *Physics in Medicine and Biology.* 2005;50(12):2887-2905.
79. Low DA, Harms WB, Mutic S, Purdy JA. A technique for the quantitative evaluation of dose distributions. *Medical Physics.* 1998;25(5):656-661.
80. Hernandez Morales D, Shan J, Liu W, et al. Automation of routine elements for spot-scanning proton patient-specific quality assurance [published online ahead of print 2018/10/20]. *Med Phys.* 2019;46(1):5-14.
81. Chi A, Chen H, Wen S, Yan H, Liao Z. Comparison of particle beam therapy and stereotactic body radiotherapy for early stage non-small cell lung cancer: A systematic review and hypothesis-generating meta-analysis. *Radiotherapy and Oncology.* 2017;123(3):346-354.
82. Lazarev S, Rosenzweig K, Samstein R, et al. Where are we with proton beam therapy for thoracic malignancies? Current status and future perspectives. *Lung Cancer.* 2021;152:157-164.
83. Simone CB, II, Bogart JA, Cabrera AR, et al. Radiation Therapy for Small Cell Lung Cancer: An ASTRO Clinical Practice Guideline. *Practical Radiation Oncology.* 2020;10(3):158-173.
84. Amini A, Verma V, Simone CB, II, et al. American Radium Society Appropriate Use Criteria for Radiation Therapy in Oligometastatic or Oligoprogressive Non-Small Cell Lung Cancer. *International Journal of Radiation Oncology, Biology, Physics.* 2022;112(2):361-375.
85. Ball D, Mai GT, Vinod S, et al. Stereotactic ablative radiotherapy versus standard radiotherapy in stage 1 non-small-cell lung cancer (TROG 09.02 CHISEL): a phase 3, open-label, randomised controlled trial. *The Lancet Oncology.* 2019;20(4):494-503.
86. Swaminath A, Parpia S, Wierzbicki M, et al. LUSTRE: A Phase III Randomized Trial of Stereotactic Body Radiotherapy (SBRT) vs. Conventionally Hypofractionated Radiotherapy (CRT) for Medically

Inoperable Stage I Non-Small Cell Lung Cancer (NSCLC). *International Journal of Radiation Oncology, Biology, Physics.* 2022;114(5):1061-1062.
87. Timmerman R, McGarry R, Yiannoutsos C, et al. Excessive Toxicity When Treating Central Tumors in a Phase II Study of Stereotactic Body Radiation Therapy for Medically Inoperable Early-Stage Lung Cancer. *Journal of Clinical Oncology.* 2006;24(30):4833-4839.
88. Bezjak A, Paulus R, Gaspar LE, et al. Safety and Efficacy of a Five-Fraction Stereotactic Body Radiotherapy Schedule for Centrally Located Non–Small-Cell Lung Cancer: NRG Oncology/RTOG 0813 Trial. *Journal of Clinical Oncology.* 2019;37(15):1316-1325.
89. Tekatli H, Haasbeek N, Dahele M, et al. Outcomes of Hypofractionated High-Dose Radiotherapy in Poor-Risk Patients with "Ultracentral" Non–Small Cell Lung Cancer. *Journal of Thoracic Oncology.* 2016;11(7):1081-1089.
90. Rwigema J-CM, Verma V, Lin L, et al. Prospective study of proton-beam radiation therapy for limited-stage small cell lung cancer. *Cancer.* 2017;123(21):4244-4251.
91. Higgins KA, O'Connell K, Liu Y, et al. National Cancer Database Analysis of Proton Versus Photon Radiation Therapy in Non-Small Cell Lung Cancer. *International Journal of Radiation Oncology, Biology, Physics.* 2017;97(1):128-137.
92. Nantavithya C, Gomez DR, Wei X, et al. Phase 2 Study of Stereotactic Body Radiation Therapy and Stereotactic Body Proton Therapy for High-Risk, Medically Inoperable, Early-Stage Non-Small Cell Lung Cancer. *International Journal of Radiation Oncology, Biology, Physics.* 2018;101(3):558-563.
93. Shi L, Tsui T, Wei J, Zhu L. Fast shading correction for cone beam CT in radiation therapy via sparse sampling on planning CT. *Medical physics.* 2017;44(5):1796-1808.
94. Oyama A, Kumagai S, Arai N, et al. Image quality improvement in cone-beam CT using the super-resolution technique. *Journal of radiation research.* 2018;59(4):501-510.
95. Ning R, Tang X, Conover D. X‐ray scatter correction algorithm for cone beam CT imaging. *Medical physics.* 2004;31(5):1195-1202.
96. Reitz I, Hesse B-M, Nill S, Tücking T, Oelfke U. Enhancement of image quality with a fast iterative scatter and beam hardening correction method for kV CBCT. *Zeitschrift für Medizinische Physik.* 2009;19(3):158-172.
97. Sonke JJ, Zijp L, Remeijer P, Van Herk M. Respiratory correlated cone beam CT. *Medical physics.* 2005;32(4):1176-1186.
98. Li T, Xing L, Munro P, et al. Four‐dimensional cone‐beam computed tomography using an on‐board imager. *Medical physics.* 2006;33(10):3825-3833.
99. Li T, Xing L. Optimizing 4D cone-beam CT acquisition protocol for external beam radiotherapy. *International Journal of Radiation Oncology\* Biology\* Physics.* 2007;67(4):1211-1219.
100. Jia X, Tian Z, Lou Y, Sonke J-J, Jiang SB. Four-dimensional cone beam CT reconstruction and enhancement using a temporal nonlocal means method. *Medical Physics.* 2012;39(9):5592-5602.
101. Yan D, Vicini F, Wong J, Martinez A. Adaptive radiation therapy. *Physics in Medicine and Biology.* 1997;42(1):123-132.
102. Albertini F, Matter M, Nenoff L, Zhang Y, Lomax A. Online daily adaptive proton therapy. *The British Journal of Radiology.* 2019;93(1107):20190594.
103. Bernatowicz K, Geets X, Barragan A, Janssens G, Souris K, Sterpin E. Feasibility of online IMPT adaptation using fast, automatic and robust dose restoration. *Physics in Medicine & Biology.* 2018;63(8):085018.
104. Borderías-Villarroel E, Taasti V, Van Elmpt W, Teruel-Rivas S, Geets X, Sterpin E. Evaluation of the clinical value of automatic online dose restoration for adaptive proton therapy of head and neck cancer. *Radiotherapy and Oncology.* 2022;170:190-197.

105. Graeff C, Constantinescu A, Luechtenborg R, Durante M, Bert C. Multigating, a 4D Optimized Beam Tracking in Scanned Ion Beam Therapy. *Technology in Cancer Research & Treatment.* 2014;13(6):497-504.
106. Yu J, Zhang X, Liao L, et al. Motion-robust intensity-modulated proton therapy for distal esophageal cancer. *Medical Physics.* 2016;43(3):1111-1118.
107. Bernatowicz K, Zhang Y, Perrin R, Weber DC, Lomax AJ. Advanced treatment planning using direct 4D optimisation for pencil-beam scanned particle therapy [published online ahead of print 2017/06/22]. *Phys Med Biol.* 2017;62(16):6595-6609.
108. Engwall E, Fredriksson A, Glimelius L. 4D robust optimization including uncertainties in time structures can reduce the interplay effect in proton pencil beam scanning radiation therapy [published online ahead of print 2018/07/18]. *Med Phys.* 2018. doi: 10.1002/mp.13094.
109. Graeff C. Motion mitigation in scanned ion beam therapy through 4D-optimization. *Physica Medica-European Journal of Medical Physics.* 2014;30(5):570-577.
110. Zhu M, Kaiser A, Mishra MV, et al. Multiple Computed Tomography Robust Optimization to Account for Random Anatomic Density Variations During Intensity Modulated Proton Therapy [published online ahead of print 2020/10/22]. *Adv Radiat Oncol.* 2020;5(5):1022-1031.
111. Kardar L, Li Y, Li X, et al. Evaluation and mitigation of the interplay effects of intensity modulated proton therapy for lung cancer in a clinical setting. *Practical radiation oncology.* 2014;4(6):e259-e268.
112. Meijers A, Knopf A-C, Crijns AP, et al. Evaluation of interplay and organ motion effects by means of 4D dose reconstruction and accumulation. *Radiotherapy and Oncology.* 2020;150:268-274.
113. den Boer E, Wulff J, Mäder U, et al. Technical Note: Investigating interplay effects in pencil beam scanning proton therapy with a 4D XCAT phantom within the RayStation treatment planning system. *Medical Physics.* 2021;48(3):1448-1455.
114. Yepes P, Randeniya S, Taddei PJ, Newhauser WD. Monte Carlo fast dose calculator for proton radiotherapy: application to a voxelized geometry representing a patient with prostate cancer. *Physics in Medicine & Biology.* 2009;54(1):N21.
115. Fix MK, Frei D, Volken W, Born EJ, Aebersold DM, Manser P. Macro Monte Carlo for dose calculation of proton beams. *Physics in Medicine and Biology.* 2013;58(7):2027-2044.
116. Wan Chan Tseung H, Ma J, Beltran C. A fast GPU-based Monte Carlo simulation of proton transport with detailed modeling of nonelastic interactions [published online ahead of print 2015/07/02]. *Med Phys.* 2015;42(6):2967-2978.
117. Jia X, Schümann J, Paganetti H, Jiang SB. GPU-based fast Monte Carlo dose calculation for proton therapy. *Physics in Medicine & Biology.* 2012;57(23):7783.
118. Schiavi A, Senzacqua M, Pioli S, et al. Fred: a GPU-accelerated fast-Monte Carlo code for rapid treatment plan recalculation in ion beam therapy. *Physics in Medicine & Biology.* 2017;62(18):7482.
119. Souris K, Lee JA, Sterpin E. Fast multipurpose Monte Carlo simulation for proton therapy using multi- and many-core CPU architectures. *Medical Physics.* 2016;43(4):1700-1712.
120. Fracchiolla F, Engwall E, Janson M, et al. Clinical validation of a GPU-based Monte Carlo dose engine of a commercial treatment planning system for pencil beam scanning proton therapy. *Physica Medica.* 2021;88:226-234.
121. Lin L, Taylor PA, Shen J, et al. NRG Oncology Survey of Monte Carlo Dose Calculation Use in US Proton Therapy Centers [published online ahead of print 20210525]. *Int J Part Ther.* 2021;8(2):73-81.
122. Favaudon V, Caplier L, Monceau V, et al. Ultrahigh dose-rate FLASH irradiation increases the differential response between normal and tumor tissue in mice. *Science Translational Medicine.* 2014;6(245).

123. Liu R, Charyyev S, Wahl N, et al. An Integrated Physical Optimization Framework for Proton Stereotactic Body Radiation Therapy FLASH Treatment Planning Allows Dose, Dose Rate, and Linear Energy Transfer Optimization Using Patient-Specific Ridge Filters. *International Journal of Radiation Oncology\*Biology\*Physics.* 2023. doi: https://doi.org/10.1016/j.ijrobp.2023.01.048.
124. Balakrishnan G, Zhao A, Sabuncu MR, Guttag J, Dalca AV. VoxelMorph: A Learning Framework for Deformable Medical Image Registration. *IEEE Transactions on Medical Imaging.* 2019;38(8):1788-1800.
125. Chen J, He Y, Frey EC, Li Y, Du Y. Vit-v-net: Vision transformer for unsupervised volumetric medical image registration. *arXiv preprint arXiv:210406468.* 2021.
126. Shi J, He Y, Kong Y, et al. Xmorpher: Full transformer for deformable medical image registration via cross attention. Paper presented at: International Conference on Medical Image Computing and Computer-Assisted Intervention2022.
127. Bai T, Wang B, Nguyen D, Jiang S. Deep dose plugin: towards real-time Monte Carlo dose calculation through a deep learning-based denoising algorithm. *Machine Learning: Science and Technology.* 2021;2(2):025033.
128. van Dijk RH, Staut N, Wolfs CJ, Verhaegen F. A novel multichannel deep learning model for fast denoising of Monte Carlo dose calculations: preclinical applications. *Physics in Medicine & Biology.* 2022;67(16):164001.
129. Pastor-Serrano O, Perkó Z. Millisecond speed deep learning based proton dose calculation with Monte Carlo accuracy. *Physics in Medicine & Biology.* 2022;67(10):105006.